\documentstyle[preprint,prb,aps]{revtex}
\tighten
\begin{document}
\draft
\preprint{MA/UC3M/6/94 \\ (Submitted to Phys.\ Rev.\ B)}

\title{Absence of localization and large dc conductance in
random superlattices with correlated disorder}

\author{Enrique Diez and Angel S\'anchez}

\address{Escuela Polit\'ecnica Superior,
Universidad Carlos III de Madrid, \\
C./ Butarque, 15, E-28911 Legan\'es, Madrid, Spain}

\author{Francisco Dom\'{\i}nguez-Adame}

\address{Departamento de F\'{\i}sica de Materiales,
Facultad de F\'{\i}sicas, Universidad Complutense,\\
E-28040 Madrid, Spain}

\maketitle

\begin{abstract}

We study how the influence of structural correlations in disordered
systems manifests itself in experimentally measurable magnitudes,
focusing on dc conductance of semiconductor superlattices with general
potential profiles.  We show that the existence of bands of extended
states in these structures gives rise to very
noticeable peaks in the finite
temperature dc conductance as the chemical potential is moved through
the bands or as the temperature is increased from zero. On the basis of
these results we discuss how dc conductance measurements can provide
information on the location and width of the bands of extended states.
Our predictions can be used to demonstrate experimentally that
structural correlations inhibit the localization effects of disorder.

\end{abstract}


\narrowtext

\section{Introduction}

A number of recent papers have proposed and provided theoretical
evidence that in disordered systems where the disorder exhibits some
kind of short range spatial correlation wave localization may be
inhibited and bands of extended states appear.\cite{Flores,Dunlap,Wu,%
Wu2,Wu3,Bovier,Wu4,Evan1,indios,Evan2,nozotro,Flores2,JPA,PRBKP,%
exciton,exciton2,bruto,exciton3} This phenomenon has been shown to arise
in a number of different contexts, like electron transport,
\cite{Flores,Dunlap,Wu,Wu2,Wu3,Bovier,Wu4,Evan1,indios,Flores2,JPA,PRBKP}
phonon transport,\cite{nozotro,bruto} exciton dynamics,%
\cite{exciton,exciton2,exciton3} or magnon propagation.\cite{Evan2} All
these theoretical analyses openly contradict the belief that
localization of {\em all} eigenstates is a general phenomenon in
one-dimensional disordered systems. Note, however that this
belief has only been rigorously proven for some {\em uncorrelated}
random systems \cite{Ziman}, and hence the existing theorems do not
apply to the above cases.  In spite of this, there is some controversy
as to the relevance of these results and the nature of the band or bands
of extended states, and delocalization by
structural correlation is still not
generally accepted.  Therefore, we undertook the task of finding {\em
experimentally measurable quantities and physically realizable systems}
that allow for a clearcut validation of the above mentioned results.  We
have already proposed experiments on mechanical models%
\cite{nozotro,bruto} but, admittedly, these may be hard to construct and
seem rather artificial or academic systems.  For this reason,
we decided to concern ourselves with a more interesting system,
namely semiconductor superlattices.

Non-periodic (quasi-periodic or disordered) semiconductor superlattices
are being studied with increasing interest in the last decade.  Firstly,
Merlin and coworkers studied Fibonacci superlattices\cite{Merlin} where
the unusual, fractal-like spectral properties give rise to very
characteristic properties.  Shortly thereafter, localization was
observed in intentionally disordered GaAs/Ga$_{1-x}$Al$_x$As
superlattices.\cite{Chomette}  This was followed by a number of studies
on disordered superlattices,\cite{Sasaki} that showed
a much larger photoluminiscence intensity than ordered lattices
\cite{Kasu} among other different features that we do not
describe here.\cite{22,Arent} Other materials like Si$_{1-x}$Ge$_x$/Si
have been shown to exhibit the same phenomenon. \cite{Wakahara} Thus,
this rather good knowledge already available as well as recent advances of
molecular beam expitaxy make these systems the ideal candidates to
propose experiments on localization or delocalization electronic
properties.

The paper is organized as follows.  In Sec.\ II we present our model and
summarize previous work of us\cite{JPA,PRBKP}
that is necessary for a better
understanding of the present paper.  The body of the paper is Sec.\ III
where we present our results on finite temperature dc conductance.  We
begin by discussing the motivation of the calculation.  Then, we proceed on
to zero temperature dc conductance which is mainly determined by the
transmission coefficient.  Most of the section is devoted to finite
temperature dc conductance.  We describe the dependence of this
magnitude on the chemical potential of the sample and on the
temperature; besides, we also study how the conductance relates to the
model parameters.  We show how the bands of extended states reveal
themselves through a well defined peak in the dc conductance.  In
addition, we also study the high temperature limit where we find a
power-law scaling of the conductance with the system size.  Finally, in
Sec.\ IV, we discuss our results and how they can be related to actual
measurements to infer the main characteristics of the bands of extended
states from experiments on superlattices.

\section{Model and Background}

\subsection{The Kronig-Penney model and its application to
superlattices}

The basis of our model is the Kronig-Penney\cite{KP} one, in which
it is assumed that the electron interaction with the underlying
one-dimensional lattice is given by a potential of the form
\begin{equation}
\label{kp}
V(x)=\sum_n\lambda_n\delta(x-x_n).
\end{equation}
This model is very general, and aside from the application to disordered
semiconductor superlattices we are going to describe, it also appears in
many other contexts like other microelectronic devices,\cite{21}
localization phenomena in liquids,\cite{20} physical properties of
layered superconductors,\cite{23} and quark tunneling in one-dimensional
nuclear models\cite{24} to name a few.  Regarding superlattices, the
choice of potential $V(x)$ given by Eq.\ (\ref{kp}) is very general as,
in principle, the superlattice potential could take many different
shapes: Square barriers, V-shaped wells, sawtooth, parabolic, etc.  In
fact, what we are doing is assuming an expression for the cell
potentials in terms of point interaction potentials.  The term point
interaction refers to any arbitrary sharply peaked potential approaching
the $\delta$ function limit (zero width and constant area).  Such
potentials are often used in a variety of physical contexts in solid
state physics,\cite{Frank} since, with limitations, they are good
candidates to replace actual, short-ranged, one-dimensional
potentials.\cite{25} Moreover, it has been recently demonstrated that
the discretized form of the Schr\"odinger equation for an arbitrary
potential in one dimension can be mapped onto a Kronig-Penney
model.\cite{BDM}  Hence the use of potential (\ref{kp}) is not a serious
restriction to simulate actual semiconductor superlattice potentials
within the envelope-function formalism.

\subsection{The Continuous Random Dimer Model.}

The version of the Kronig-Penney model we are interested in is the
so-called Continuous Random Dimer Model (CRDM), which was first
introduced by us in Refs.\ \onlinecite{JPA} and \onlinecite{PRBKP} as a
realistic theoretical scenario where delocalization effects have
dramatic consequences.  The model is defined by particularizing Eq.\
(\ref{kp}) as follows: First, we choose $\lambda_n>0$; the extension of
the results to the $\lambda_n<0$ case is straightforward, although
the choice of the sign is irrelevant for the superlattice
application as $\lambda_n$ may be always taken as positive by a suitable
assignation of the $\delta$ function to superlattice blocks.  Second, we
take the positions of the $\delta$ potentials to be regularly spaced,
i.e., $x_n=n$.  Third and most important, we introduce a paired
correlated disorder which implies that $\lambda_n$ takes only on two
values, $\lambda$ and $\lambda'$, with the additional constraint that
$\lambda'$ appears only in pairs of neighboring sites (dimer).  The
corresponding Schr\"odinger equation is then ($\hbar=m=1$)
\begin{equation}
\label{Schr}
\left[-{d^2\phantom{x}\over dx^2} + \sum_n \lambda_n
\delta(x-n)\right] \psi(x) = E\>\psi(x).
\end{equation}

In Ref.\ \onlinecite{PRBKP} we developed a generalized Poincar\'e map
formalism that allows to map general one-dimensional Schr\"odinger
equations onto discrete equations exactly, for any potential allowed in
quantum mechanics.  In particular, its application to Eq.\ (\ref{Schr})
is quite simple.  For the sake of brevity, we only quote here the final
result, and refer the reader to Sec.\ II of Ref.\ \onlinecite{PRBKP} for
the details.  Equation (\ref{Schr}) is {\em exactly} equivalent to
\begin{equation}
\label{map}
\psi_{n+1}+\psi_{n-1} = \left[2\cos q + {\lambda_n\over q} \sin q
\right] \psi_n,
\end{equation}
where we have put $\psi_n\equiv\psi(x=n)$ and $q\equiv\sqrt{E}$.

{}From the above equation, we can see that there are an infinite number
of resonant energies for which the reflection coefficient of a {\em
single} dimer vanishes.  Indeed, taking into account that the condition
for an electron to move in the perfect lattice [$\lambda_n =\lambda$ for
all $n$ in Eq.\ (\ref{map})] is
\begin{equation}
\label{condition}
\left|\cos q + {\lambda\over 2q}\sin q\right| \leq 1,
\end{equation}
we have a first restriction on the allowed values of energy.  Further,
introducing a single dimer occupying sites $n=0$ and $n=1$ and
eliminating $\psi_0$ and $\psi_1$ we obtain
\begin{equation}
\label{cond2}
-\psi_2 = (\omega+\omega'-\omega'^2\omega)\psi_{-1} -
(1-\omega'^2)\psi_{-2},
\end{equation}
with $\omega\equiv2\cos q+(\lambda/q) \sin q$ and $\omega'$ the same
exchanging $\lambda$ by $\lambda'$.  It is evident from this expression
that if $\omega'=0$ we recover the equation for the perfect lattice with
sites $n=0$ and $n=1$ suppressed except for an irrelevant phase factor
$\pi$.  This means that at the particular values $q_r$ such that
$\omega'=0$ the reflection coefficient of the dimer vanishes.  Such
condition and the perfect lattice one (\ref{condition}) yield the two
equations determining the resonances $E_r=q_r^2$
\begin{mathletters}
\begin{eqnarray}
\label{1}
|\cos q_r| & \leq & {\lambda'\over |\lambda-\lambda'|}, \\
\label{2}
-{2\over\lambda'} & = & {\tan q_r\over q_r},
\end{eqnarray}
\end{mathletters}
which is our final result.  Restricting ourselves to the range $\lambda
\leq 2\lambda'$ Eq.\ (\ref{1}) is trivially satisfied.  Then, Eq.\
(\ref{2}) has an infinite number of solutions, one in every interval
$[(2n-1)\pi/2,(2n+1)\pi/2],\>n=1,2,\ldots$ leading to infinite energy
values for which the reflection coefficient of a single dimer vanishes.

\subsection{Properties of the model}

Of course, the above result does not imply anything about extended
states in a CRDM with a {\em finite} density of dimers, and it is
necessary to study that problem separately.  This we carried out in
Refs.\ \onlinecite{JPA} and \onlinecite{PRBKP} by means of numerical
evaluation of exact expressions obtained via transfer matrix techniques
for the relevant quantities: Transmission coefficient, Landauer
resistance, Lyapunov coefficient, and density of states.  The behavior
of all these quantities, combined with multifractal and inverse
participation ratio analyses, allowed us to establish on firm grounds
that the single dimer resonances survive in the CRDM and, moreover, that
they give rise to bands of finite width of truly extended states.  The
interested reader may find a thorough report in Ref.\
\onlinecite{PRBKP}.  Here we will only comment on one of these
magnitudes, namely the transmission coefficient, which is the starting
point for our computations of finite temperature dc conductance.

An example of the behavior of the transmission coefficient around one of
the resonant energies is shown in Fig.\ \ref{trans} for a dimer
concentration $c=0.5$ ($c$ is defined as the ratio between the number of
$\lambda'$ and the total number of $\delta$'s in the lattice).  We
stress that, in spite of the fact that the plot corresponds to an
average over 100 realizations of the CRDM, the transmission coefficient
for typical realizations behaves in the same way, although noisier.
Thus, the only effect of averaging is to smooth out particular features
of realizations keeping only the main common characteristic, i.e., the
wide transmission peak.  This is the property we want to highlight:
Close to single dimer resonances (in the case of Fig.\ \ref{trans}, the
first one, which occurs at $E_r=3.7626\ldots$ for the chosen parameters
$\lambda=1.0$, $\lambda'=1.5$), there is an interval of energies that
shows also very good transmission properties, similar to those of the
resonant energy.  Most important, such interval has always a finite
width, for all values of dimer concentration, $\lambda$ and $\lambda'$
(provided they satisfy the above conditions), or number of sites in the
lattice.  The peak width depends on the order of the resonance (the
higher the resonance the wider the band of states with transmission
coefficient close to unity) and the concentration of dimers (the larger
the concentration, the narrower the peak, being always of finite width
as already stated).  Fig.\ \ref{trans} also shows an analytical fit to
the shape of the transmission coefficient dependence on the energy,
which will be used below.  The parametrization used in the best fit
close to $E_r$ is $\tau(E)=m(E)+g(E)$, where

\begin{equation}
\label{moyal}
m(E)= m_0 \exp \left(-\frac {v + e^{-v}}{2} \right) \nonumber
\end{equation}
being the Moyal\cite{Moyal} function with $v=(E-E_r)/s$, and
\begin{equation}
\label{gaussi}
g(E)= g_0 \exp \left( -\frac {(E-E_r)^2}{2\sigma^2} \right) \nonumber
\end{equation}
being the usual gaussian function.  The parameters that fit data in
Fig.\ \ref{trans} are $m_0=0.65$, $s=0.16$, $g_0=0.70$, and
$\sigma=0.26$.  As can be seen from Fig.\ \ref{trans}, the fitting
reflects the asymmetry of the peak and corresponds very well to the
average transmission coefficient.  Of the above constants, the most
relevant one is probably the variance of the gaussian; we will come back
to this parameter and its relevance below.

\section{Finite temperature dc conductance}

\subsection{Motivation: Characteristics to determine}

So far, we have summarized the main properties of the CRDM, which can be
found in full detail in Ref.\ \onlinecite{PRBKP}.  The crucial
conclusion of those previous studies has been already mentioned: There
are bands (an infinite number of them) of truly extended states in the
CRDM in spite of the disorder.  We have provided enough theoretical
evidence and we can be quite sure of the correctness of that statement.
The most important point, however, regards {\em applications} of this
result, and this inmediately implies two questions: First, are these
bands of extended states {\em experimentally} measurable?  Admittedly,
if the extended states we have predicted are not seen in actual physical
systems, the question as to their true extended nature becomes
irrelevant.  Second, do these extended states serve as the basis for new
applications or devices?  Hopefully, the answer to this second question
would be yes {\em provided} the answer to the first question was also
yes.  This may be easily understood if we think that the transport
properties of such microstructures would depend strongly on the value of
the incoming energy, and therefore they could serve as ``filters'' of
unwanted energies (in fact, we have proposed similar applications in
mechanical devices based on the same ideas \cite{bruto}).  It is then
clear than the crux of the problem is the first question, and that would
be the one we will try to answer in the remainder of the paper.
Specifically, we will devote ourselves to show how the position of the
peak and its width may be determined from finite temperature dc
conductance measurements, and how experiments relate to the dimer
concentration.

\subsection{Zero temperature dc conductance}

In this subsection, we discuss electron propagation at very low
temperatures through a disordered superlattice with one of the two
alternating types of constituents subject to the constraint of the CRDM.
We note at this point that the $\delta$ function may represent the joint
potential of several layers, e.g., a block GaAs-GaAlAs-GaAs giving rise
to a square potential barrier.  Different choices of the blocks are then
associated to the two different values of $\lambda$ and $\lambda'$.  We
will term these superlattices correlated disordered superlattices
(CDSL).  In general, two main factors must be taken into account when
dealing with vertical transport through a CDSL.  On the one hand, since
this is essentially a quantum phenomenon, we must consider systems with
strong coupling between adjacent blocks, but in our model this is not a
problem since the distance between $\delta$ functions does not play any
r\^ole aside from fixing the resonant energies.  On the other hand, we
are neglecting electron-phonon scattering effects which tend to disrupt
coherent quantum transport.  These effects crucially depend on the
sample temperature, so it may be confidently expected that their
influence can be neglected at very low temperatures.  Besides,
superlattices used for experiments have periods in the range from one
monolayer to several nanometers.
\cite{Chomette,Sasaki,Kasu,22,Arent,Wakahara} A short spacing between
layers also contributes to reduce the scattering by phonons, and
therefore, short-period superlattices could be useful for the work we
propose up to higher temperatures, because as we have seen the period
length is not very relevant.  Hence, a physical realization of our model
is possible and the measurements should be comparable to our predictions
in a wide range of temperatures.

The electrical conductance at zero temperature can be obtained
straightforwardly from the well-known dimensionless single-channel
Landauer formula\cite{Landauer}
\begin{equation}
\label{landauer}
\kappa_0(E)={\tau(E)\over1-\tau(E)}.
\end{equation}
We have already shown the behavior of the transmission coefficient as a
function of the energy in the previous section, as obtained by means of
the transfer matrix formalism.\cite{PRBKP} The calculation of the
Landauer conductance is then straightforward for any value of the
parameters using the same approach.  A typical example of the results is
shown in Fig.\ \ref{kcero} for the same values of the parameters as in
Fig.\ \ref{trans}.  For a single realization, Fig.\ \ref{kcero}(a) shows
that the detailed structure of the energy spectrum naturally determines
the finer details of the conductance pattern at zero temperature.  Thus
the noisy aspect of the curve.  However, by comparing it to the average
plotted in Fig.\ \ref{kcero}(b), we realize that the result is the very
close to the single realization one (as we mentioned when discussing the
transmission coefficient), except for the fact that some particular,
realization-dependent conductance spikes are suppressed.
Note that the computation involves the ratio of the reflection to the
transmission coefficient and this quotient enlarges fluctuations
considerably, thus the noisy aspect of Fig.\ \ref{kcero}(b). Obtaining
an smoothing as in Fig.\ \ref{trans} would involve averaging over
many more realizations. This is very
important and we will take advantage of this fact when we discuss how to
measure the width of the extended states band, but we can already
advance that this magnitude is of order of the width of the peak of
$\kappa_0$ and that it can be determined from a single realization,
i.e., a single superlattice.  Finally, we note that the behavior
reported here for the first resonance is equally verified for the
subsequent ones, so the discussion is not restricted to this first band
which should be merely taken as an example.

\subsection{Finite temperature dc conductance}

We now proceed to compute the electrical conductance at any temperature.
We have already computed this magnitude and its relation to the energy
spectrum for Fibonacci superlattices in previous works.\cite{Fibo,moco}
Those works gave results in agreement with known facts about this
kind of superlattices (see discussions in Ref.\ \onlinecite{Fibo}). We can
then be confident that our calculation will also be relevant in the physical
context we are dealing with. The idea in including temperature effects
is twofold. First, we then have a control parameter that can be varied
at will and correspondingly we can have a whole set of measurements.
Second, the previous results on very low temperature conductance have
a limited range of validity and in order to compare with experiments
temperature should be included in our results.

The dimensionless finite temperature conductance can be obtained through
the following expression, earlier discussed in detail by Engquist and
Anderson\cite{28}
\begin{equation}
\label{kt}
\kappa(T,\mu)= {\int \left(-{\partial n\over \partial E}
\right) \tau(E)dE \over \int \left(-{\partial n\over \partial E}
\right) [1-\tau(E)]}dE,
\end{equation}
where integrations are extended over the allowed bands, $\tau(E)$ is the
transmission coefficient, $n$ is the Fermi-Dirac distribution, and $\mu$
denotes the chemical potential of the sample.  We have calculated
expression (\ref{kt}) numerically using the transmission coefficient as
input.  We discuss separately the cases of low temperatures and the high
temperature limit in the following.

\subsubsection{Low temperatures}

A global view of the results is presented in Fig.\ \ref{3D}.  Again,
averages smooth out the realization dependent features and preserve the
common structure, namely the clear peak in the conductance around the
resonant energy.  We also show for comparison what is obtained in the
case when we remove the dimer constraint, i.e., for a purely random
superlattice.  Taking into account the largely different scales between
Figs.\ \ref{3D} (a) and (c), the uncorrelated
random superlattice conductivity is
very low for all energies.  The small features appearing in the plot are
specific of the chosen realization and when one takes averages the
final result is a flat, zero plot.  We then prove that there should be
an enormous increase in conductivity, clearly noticeable through
experiments when the dimer constraint is satisfied. We note in passing
that the result for the pure random system confirms the validoty of
our procedure.

In those plots, the results discussed for zero temperature are also
included, and it is clear that for not so large temperatures the system
exhibit a conductance behavior that reproduces the transmission coefficient
of the first band, around which the figure is centered.  As we start
increasing temperature, the peak lowers and widens, and it is already
difficult to appreciate for values of $kT$ around 0.5 (around a 5\% of
the perfect lattice bandwith).  It is remarkable that in this high
temperature region the conductance is clearly non-zero, to be compared
to that of the pure random system.  It is not difficult to understand
why this is so.  For low temperatures, the derivative of the Fermi-Dirac
function is very peaked around the chemical potential.  Therefore, only
when the chemical potential is close to the band of extended states,
that is to say, close to the resonance, there will be positive
contributions to the conductance, and chemical potentials far from the
resonance will show zero conductance.  As temperature is increased, the
derivative of the Fermi-Dirac function becomes wider, and consequently
it is not necessary to choose a chemical potential close to the
resonance; even if it is placed far from it the integrals will include
the contribution of the extended states.  On the contrary, the peak
height decreases because previously, for chemical potentials in the
band, the localized states outside were not present in the integration,
whereas for larger temperatures they contribute in a negative fashion to
the conductance properties, and they are weighted more in the
integration.  The behavior we show in Fig.\ \ref{3D} coincides then with
the intuitive expectations.

At this point, we recall the analytical approximation we presented in
Fig.\ \ref{trans} for the transmission coefficient.  An evident way to
check whether this magnitude actually behaves in average in such a
smooth manner is to use the fitted expression in Eq.\ (\ref{kt}) and
compute again the conductance.  As is depicted in Fig.\ \ref{teo}, the
agreement is very good between both procedures.  We can then assume that
our fitting is a correct description of the transmission coefficient
dependence on the temperature even for each realization, provided that
we are not interested in the particular, noisy characteristics of them.
This is so because Fig.\ \ref{3D} (b) has been computed by generating
realizations of the model, computing the transmission coefficient and
the conductance, and {\em after that} averaging this last quantity,
whereas the theoretical calculation in Fig.\ \ref{teo} uses the analytic
expression for the transmission coefficient only one, and no averages
are involved.  That is why we may conclude that the analytic expression
can be used for a typical realization.  Comparison with transmission
coefficients plots for a single realization is also satisfactory.
Finally, we point out that the main contribution to the conductance
comes from the extended states close to the resonance.  This we checked
by using a simple parabolic fit to the transmission coefficient,
neglecting the tails.  The result is again very similar to Fig.\
\ref{3D}, reinforcing our previous conclusion.

\subsubsection{High temperature limit}

Figure \ref{3D} indicates that the conductance curve rapidly saturate
towards a value $\kappa_{\infty}$, independent of the chemical
potential, when the temperature reaches a value of the order of 0.5
$kT$.  The reason why the asymptotic dependence on the temperature is
independent of $\mu$ is easy to understand, and we have already
discussed it in Ref.\ \onlinecite{Fibo}.  The idea goes as follows:
Assume we are in the high temperature regime.  In this regime, all
electrons contribute to vertical transport.  Now, define $\epsilon\equiv
E-\mu$; for high temperatures $\beta\epsilon<\!< 1$ and we can expand
the Fermi-Dirac derivative as
\begin{equation}
\label{expan}
-{\partial n\over\partial E}\simeq {1\over 4} + O[(\beta\epsilon)^2],
\end{equation}
which inserted into the general formula (\ref{kt}) yields
\begin{equation}
\label{hight}
\kappa_{\infty}\simeq {\int \tau(E)dE \over \int [1-\tau(E)]}dE,
\end{equation}
an expression where the chemical potential has disappeared, and the only
dependence is on the number of $\delta$ functions, their strengths, and
their concentrations, all these quantities entering through the
transmission coefficient.  In fact, we have checked that
$\kappa_{\infty}$ scales with the number of scatterers as a power law,
with exponent depending on the dimer concentration as we show in Fig.\
\ref{scaling}.  Similar results are obtained by changing the strengths
of the $\delta$ functions.  On the other hand, from these same
reasonings it can be induced that the vanishing of the high temperature
conductance of the pure random system is due to the fact that the
transmission coefficient is always close to zero, and there are no
extended states.  Once again we see that the dimer structure gives rise
to largely new features as compared to the random one.

\section{Discussion and conclusions}

After reporting all our study of the finite temperature dc conductance
of CDSL's, we are now in the best position to discuss what are the means
to obtain the characteristics of the band of extended states from
experimental measurements.  We mentioned in the motivation that we
intended to find the position of the bands, for instance the first one
and their width.  We begin with the system that should be used: We
believe that any of the experimental setups used in previous works,
short period GaAs/GaAlAs (Refs.\
\onlinecite{Chomette,Sasaki,Kasu,22,Arent}) or SiGe/Si (Ref.\
\onlinecite{Wakahara}) are suitable devices for the kind of
measurementes we proposed, provided that they are built with the dimer
constraint.  Electronic transport through the so built CDSL can be
measured either by techniques that employ magnetic or electric
fields,\cite{tec1} or by all-optical procedures,\cite{tec2} or by a
combination of both that helps avoid the intrinsic experimental problems
of each of them.

Let us now turn to the measurements themselves.  The first quantity we
have to determine is the position of the extended band of states.  The
way to do that is to prepare several superlattices with different
chemical potentials.  This could be achieved by varying doping
concentration or pressure.  These samples would have different chemical
potential, but as far as the nature of the layers forming the dimer is
not changed, the position of the bands must be the same.  Therefore, the
particular sample for which a maximum of the conductivity were reached
would be that with the chemical potential closer to the required band.
Even if the first band is well below the Fermi level, as there are
infinite other bands, some of them would be reached and a rapid increase
in conductivity should be noticeable.  Further, higher order bands are
wider, so they should be even easier to detect.  The question remains as
to what is the appropriate range of temperatures to look at, because if
temperature is too high no maximum should be detected.  As we already
mentioned, this limit is reached at about $kT$ of order the of 5\% of
the perfect lattice bandwidth.  In typical superlattices this width is
close to $100\,$meV (say).  Thus high temperature limit means that
$kT\sim 5\,$meV, that is, $T$ around liquid nitrogen temperature.
Therefore, the marked peak in the dc conductance should be clearly
observable at temperatures close to 5--10\,K in most superlattices with
different values of the chemical potential.  The practical implications
of this result is twofold.  First, the range of temperature is
physically realizable and second, electron-phonon interaction can be
neglected, as we have assumed.

So, we may suppose that we have hit a band of extended states, and that
we know approximately the location of the center.  The next step is to
perform a number of measurements, for some values of the chemical
potential at different temperatures.  If the chemical potential does not
need to be varied much the use of CDSL's with different dimer
concentrations could be considered as well.  If we were able to measure
the conductance at zero temperature, we would have a portrait of the
transmission coefficient itself.  However, this is not possible, but
what we can always do is measure as close to zero as to have an idea of
what is the shape of the transmission coefficient.  A more quantitative
way to do this is the following: Take a series of measurements of the
conductance for different values of temperature.  For each of the so
obtained profiles (actually sections of Fig.\ \ref{3D}) one can compute
its width by a number of means (and also depending on the definition of
width itself).  For instance, what one can do is to fit gaussians to the
experimental profiles, if it is not desired to use the more
sophisticated function mentioned in Sec.\ II. That would provide the
width $\Delta$ as a function of $T$.  We carried out this in our model
and the result is shown in Fig.\ \ref{sigma}.  In this plot we may note
that the width behaves as  $\Delta (T)=a_0+a_1 T+ a_2 T^2$, with
$a_0\sim 0.05$, $a_1\sim 1$ and $a_2\sim 1$.  The value of the width of
the band of extended states is then simply $a_0$ (i.e., the fitting
evaluated at $T=0$).  We have thus provided a means to estimate the
width of this band for CDSL's.

Finally, to ensure consistency of all the procedure, and also to help
choose the better regime to work on, we can also examine how the peak
width depends on the concentration of dimers and also how does it behave
with temperature for a given concentration.  The results are shown in
Figs.\ \ref{con1} and \ref{con2}.  Fig.\ \ref{con1} exhibits a
dependence of the width $\Delta$ on the dimer concentration basically as
$c^{-1}$.  This makes sense, because for concentration almost zero the
superlattice would be practically perfect, and the good properties of
the dimers would leave most states unscattered.  If we now look at the
temperature where the maximum of the conductance is achieved, we see
that it moves towards higher temperatures (see Fig.\ \ref{con2}), with a
functional form which is again roughly $c^{-1}$.  This may facilitate
working on higher temperatures if needed.

In conclusion, we have studied finite temperature conductance of the
CRDM and show how this study leads to specific predictions that may be
measured on actual superlattices.  If performed, those experiments would
validate (or discard) all the recent claims that correlation induces the
appearance of bands of extended states in spite of the localization
effects of disorder.  Aside from our suggestions above of already
studied quantum well superlattices, it is most interesting to note tha
recent results on a single Si cell with double $\delta$ doping
\cite{hala} have been reported, and they exhibit a large increase of
electron mobility in this kind of structure as compared to single or
homogeneously doped structures.  Although our analysis may not apply
straightforwardly to this measurement, it is tempting to suggest that at
the roots of the behavior may be the dimer resonance effect, at least
partially.  On the other hand, $\delta$ doped structures might be even
more suitable to this kind of experiments and fit better the theoretical
model we have been discussing.  We hope that this work stimulates
experimental efforts in this direction.

\acknowledgments

It is with great pleasure that we thank collaboration and illuminating
conversations with Enrique Maci\'a.  A.\ S.\ is also thankful to Alan
Bishop for warm hospitality at Los Alamos National Laboratory where this
paper was written in part.  Work at Legan\'es is supported by the
Direcci\'on General de Investigaci\'on Cient\'\i fica y T\'ecnica
(Spain) through project PB92-0378, and by the European Union Human
Capital and Mobility Programme through contract ERBCHRXCT930413.  Work
at Madrid is supported by Universidad Complutense through project
PR161/93-4811.

\begin{figure}
\caption{Transmission coefficient for the CRDM with a dimer
concentration $c=0.5$.  The $\delta$ function strengths are
$\lambda=1,\>\lambda'=1.5$.  Shown is an average over 100 realizations.
Every realization consists of 15 000 scatterers.  The first allowed band
in the perfect lattice is $[0.921,9.870]$.  The solid line is an
analytical fit using Moyal and gaussian functions (see text for
details).}
\label{trans}
\end{figure}

\begin{figure}
\caption{Landauer conductance at zero temperature for a CDSL with the
same parameters as in Fig.\ \protect\ref{trans}.  (a) A single
realization and (b) an average over 100 realizations.}
\label{kcero}
\end{figure}

\begin{figure}
\caption{Finite temperature conductance as a function of temperature and
chemical potential for a CDSL with the same parameters as in Fig.\
\protect\ref{trans}.  (a) A single realization, (b) an average over 100
realizations, and (c) a single realization without the dimer constraint,
i.e., pure randomly disordered lattice.  Note the very different
vertical scales.}
\label{3D}
\end{figure}

\begin{figure}
\caption{Finite temperature conductance obtained using the fitting with
Moyal and gaussian functions in Fig.\ \protect \ref{trans} in Eq.\
(\protect\ref{kt}).}
\label{teo}
\end{figure}

\begin{figure}
\caption{Dependence of the high temperature conductance on on the
sample length for several values of the concentration showing their
power-law behavior. Solid lines are least squares fits with slopes
of order of $-0.85$.}
\label{scaling}
\end{figure}

\begin{figure}
\caption{Dependence of the width of the conductance peak on the
temperature as given by gaussian fits.  System parameters are as in
Fig.\ \protect \ref{trans}.}
\label{sigma}
\end{figure}

\begin{figure}
\caption{Dependence of the width of the conductance peak at $kT=0.02$ on
the dimer concentration as given by gaussian fits.  System parameters
are as in Fig.\ \protect \ref{trans}.}
\label{con1}
\end{figure}

\begin{figure}
\caption{Conductance as a function of temperature for different
concentrations.  Chemical potential is always placed at the middle of
the band of extended states.}
\label{con2}
\end{figure}

\end{document}